\documentclass[12pt,preprint]{aastex}

\shorttitle{The dust heating mechanism in radio-loud AGN}
\shortauthors{Tadhunter et al.}

\begin{document}


\title{The heating mechanism for the warm/cool dust in powerful, radio-loud AGN}


\author{C. Tadhunter, D. Dicken, J. Holt, K. Inskip}
\affil{Department of Physics \& Astronomy, University of Sheffield, Hounsfield Road, Sheffield S3
7RH, UK}
\email{c.tadhunter@sheffield.ac.uk,d.dicken@sheffield.ac.uk,j.holt@sheffield.ac.uk,\\k.inskip@sheffield.ac.uk}

\author{R. Morganti}
\affil{ASTRON, PO Box 2, 7990 AA Dwingeloo, The Netherlands}
\email{morganti@astron.nl}

\author{D. Axon, C. Buchanan}
\affil{Department of Physics \& Astronomy, Rochester Institute of Technology, 84 Lomb 
Memorial Drive, Rochester, NY 14623, USA} 
\email{djasps@rit.edu,clbsps@rit.edu}

\author{R. Gonz\'{a}lez Delgado}
\affil{Instituto de Astrofisica de Andalucia (CSIC), Apdto. 3004, 18080 Granada, Spain} 
\email{rosa@iaa.es}

\author{P. Barthel}
\affil{Kapteyn Astronomical Institute, University of Groningen, PO Box 800, 9700 AV 
Groningen, The Netherlands} 
\email{pdb@astro.rug.nl}
\and

\author{I. van Bemmel}
\affil{Leiden Observatory, PO Box 9513, 2300 RA Leiden, The Netherlands}
\email{bemmel@strw.leidenuniv.nl}
\begin{abstract}
The uncertainty surrounding the nature of the heating mechanism for the dust 
that emits at mid- to far-IR (MFIR) wavelengths in active galaxies limits our 
understanding of the links between active galactic nuclei (AGN) and galaxy 
evolution, as well as our ability to interpret the prodigious infrared and sub-mm 
emission of some of the most distant galaxies in the Universe. Here we report 
deep {\it Spitzer}  
observations of a complete sample of powerful, intermediate redshift ($0.05 < z < 0.7$)  
radio galaxies and quasars. We show 
that AGN power, as traced by [OIII]$\lambda$5007 emission, is strongly correlated with both the mid-IR
(24$\micron$) 
and the far-IR (70$\micron$) luminosities, however, with increased scatter in the 70$\micron$ 
correlation. A major cause of this increased scatter is a group of objects that falls 
above the main correlation and 
displays evidence for prodigious recent star formation activity at optical wavelengths,
along with relatively cool MFIR colours. 
These results provide evidence that illumination by the AGN is the 
{\it primary} heating mechanism for the dust emitting at both 24 and 70$\micron$, with 
starbursts dominating the heating of the cool dust in only 20 -- 30\% of objects. This 
implies that powerful AGN are not always accompanied by the type of luminous starbursts 
that are characteristic of the peak of activity in major gas-rich mergers. 
\end{abstract}





\section{Introduction}

In hierarchical galaxy evolution scenarios it is predicted that the gas flows associated 
with the galaxy mergers that build massive galaxies will trigger both starbursts and 
AGN activity \citep{kauffmann00,dimatteo05}. As the mergers proceed, the outflows driven 
by the AGN eventually 
become powerful enough to limit both the star formation in the host galaxies and any 
further growth of the super-massive black holes \citep{dimatteo05}. In this context, there 
is clearly an 
interest in studying the co-evolution of AGN and their host galaxies. However, from an 
observational perspective, separating those features of galaxies that are associated with 
AGN, from those that are associated with star formation activity, has often proved 
problematic. For example, deep surveys at sub-mm wavelengths have been successful at 
detecting the redshifted far-IR emission from cool dust components in high redshift 
galaxies that show evidence for AGN activity at X-ray \citep{alexander05}, 
optical \citep{priddey03} and radio \citep{archibald01,willott02} 
wavelengths, but the interpretation of these results in terms of star formation activity 
remains controversial because of uncertainties surrounding the heating mechanism for 
the cool dust (e.g. Willott et al. 2002). 

Although it is generally accepted that the warm dust emitting the mid-IR (3 --
30$\mu$m) continuum is situated relatively close to the AGN and heated by direct AGN 
illumination \citep{pier92,vanbemmel03,rowan95}, the heating mechanism for the cooler, far-IR 
(30 -- 150$\mu$m) emitting dust 
is less certain because the distribution of the cool dust is unknown. There is plenty of 
observational evidence that starbursts can produce prodigious far-IR and sub-mm 
radiation. However, it is also possible to model the far-IR spectral energy distributions 
(SEDs) solely in terms of AGN heating, provided that sufficient AGN energy is allowed to 
escape to relatively large radii in order to heat a significant mass of dust to the requisite 
cool temperatures \citep{nenkova02,vanbemmel03}. 

A promising alternative to SED modelling is to use a statistical approach: 
correlating the MFIR continuum properties with information about the level of both 
AGN and starburst activity derived from observations at other wavelengths. However, 
this approach has been hampered in the past by the low sensitivity of the available far-
IR satellites, and the well-known biases that can occur in luminosity-luminosity plots of 
incomplete flux-limited samples. For example, the {\it Infrared Astronomical Satellite} (IRAS) 
detected fewer 
than 30\% of powerful 3C radio galaxies at MFIR wavelengths \citep{impey93,heckman94},
and the detection rate did not improve substantially in observations made by the
{\it Infrared Space Observatory} (ISO).  
Therefore, while 
some previous studies hinted at correlations between AGN and MFIR activity
\citep{impey93,heckman94,hes95,haas03}, 
none were definitive because of the incompleteness of the detections at far-IR 
wavelengths. Moreover, based on IRAS results, 
it was noted that some of the radio-loud AGN with the most luminous far-IR 
emission are associated with prodigious recent star formation activity detected at 
optical wavelengths, thus supporting the alternative starburst heating 
mechanism \citep{hes95,tad02,wills02,wills04}. 

The launch of the {\it Spitzer Space Telescope} \citep{werner04}, with its orders of magnitude 
improved sensitivity at MFIR wavelengths compared with previous satellites, has 
substantially enhanced our ability to make statistical studies of complete samples of 
distant AGN. In this paper we report results from a deep survey with the {\it Spitzer} MIPS 
instrument \citep{rieke04} of a complete sample of intermediate redshift  radio galaxies.
These results have a direct bearing on our understanding of the dominant heating mechanism(s)
for the warm/cool dust in AGN.

\section{Sample Selection and Observations}

Our sample comprises all radio galaxies and steep-spectrum radio quasars
with intermediate redshifts ($0.05 < z < 0.7$)
from the sample of southern 2Jy radio sources ($S_{2.7GHz} > 2.0$~Jy) described
in \cite{tad93}, with the addition of PKS0345+07 which has since proved to 
fulfill the same selection criteria \citep{diserego04}. This 2Jy sample (47 objects
in total) is unique 
in the sense that deep optical spectra exist for all the sample objects which 
can be used to derive accurate 
emission line luminosities \citep{tad93,tad98},
and search for signs of optical starburst activity \citep{tad02,wills04}.

For the majority of objects in the sample (42) we made deep {\it Spitzer} observations with the
MIPS instrument at 24 and 70$\mu$m as part of a programme dedicated to understanding the dust
heating mechanism, with typical exposure times of 92 -- 180s at 24$\mu$m and 
231 -- 545s at 70$\mu$m (depending on the brighteness). For 4
further objects we used MIPS observations already present in the {\it Spitzer} archive, and
for the remaining object --- PKS1549-79 --- we used 25 and 60$\mu$m flux measurements
obtained by IRAS. The {\it Spitzer} data were reduced using the MOPEX software package, with additional
median filtering performed using contributed software. Flux measurements were made using
the aperture photometry option in the Starlink Gaia package, with typical aperture sizes
of 12 -- 30 arcseconds and 25 -- 50 arcseconds at 24$\mu$m and 70$\mu$m respectively. 
In all cases appropriate corrections for aperture losses were made
using empirically-determined curves of growth determined from measurements of the 
brighter sources
in our sample. Our 
Spitzer observations detect 100\% of the sample at 24$\mu$m and 89\% of the sample at 
70$\mu$m. Typical flux uncertainties range from $\sim$30\% in the case of the
faintest sources in our sample, to $\sim$10 -- 20\% for the brightest. 
A more detailed presentation of the data and results will be
made in a forthcoming paper (Dicken et al., in preparation).

The continuum fluxes were converted to luminosities using $H_0 = 71$~km
s$^{-1}$ Mpc$^{-1}$, $\Omega_{m}=0.27$ and $\Omega_{\lambda}=0.73$, 
along with spectral indices derived from the measured
F(70)/F(24) flux ratios.

\section{Results}





Previous studies have correlated the MFIR properties with the radio luminosities, 
which are related to the mechanical powers of the relativistic jet components
(e.g. Hes et al. 1995; Shi et al. 2005). In 
contrast, we prefer to investigate correlations with the [OIII]$\lambda$5007 emission line 
luminosities ($L_{[OIII]}$), which provide a more direct indication of the intrinsic radiative powers of 
the illuminating AGN \citep{rawlings91,tad98,simpson98}. The main results are shown in Figures 1 and 2, which 
demonstrate that strong correlations exist between $L_{[OIII]}$ and both the mid-IR 
(24$\mu$m) and far-IR (70$\mu$m) monochromatic luminosities over four orders of magnitude 
in optical emission line luminosity. Restricting our analysis to redshifts $z>0.06$, in order 
to avoid most of the low luminosity objects in our sample with upper limits on their [OIII] 
luminosities, a Spearman rank correlation analysis shows that both correlations are 
highly significant (see Table 1). By fitting straight lines 
to the correlations in log-log space we find that their power-law slopes are consistent 
within the uncertainties: $L_{24\mu m} \propto L_{[OIII]}^{0.81\pm0.07}$  and 
$L_{70\mu m} \propto L_{[OIII]}^{0.94\pm0.1}$  for the full $z>0.06$ sample 
($n=39$); and  $L_{24\mu m} \propto L_{[OIII]}^{0.74\pm0.05}$ and 
$L_{70\mu m} \propto L_{[OIII]}^{0.81\pm0.08}$  if we exclude the objects with 
evidence for optical starburst activity ($n=32$; see below). The uncertainties in the slopes  
for the correlations have been estimated using a 
bootstrap re-sampling technique\footnote{We used 500 cycles in the bootstrap. In the case of the
70$\mu$m correlation we handled the four objects with 70$\mu$m upper limits as follows: for each 
cycle  we generated a 70$\mu$m luminosity for each of the upper limits
by multiplying the measured 24$\mu$m luminosity
by a value for the 70$\mu$m/24$\mu$m ratio drawn at random from the distribution of
such ratios measured for the sample as a whole}. One object in the $z>0.06$ sample used for 
the correlation analysis
has only an upper limit on its [OIII] luminosity. For this object we used the upper limit,
rather than a measured luminosity, in the correlation analysis.

Despite the similarities between the two correlations shown in Figures 1 and 2, the 
70$\mu$m correlation shows a larger scatter (see the final column in Table 1). Part of the reason for this larger scatter 
becomes clear when the evidence for optical starburst activity is considered. Careful 
spectral synthesis modelling of high quality optical spectra for the 2Jy sample, taking full account of 
AGN-related continuum components (see Tadhunter et al. 2002, 2005 for details), has allowed us to identify the objects that show 
strong evidence for recent starburst activity in their early-type host galaxies (highlighted 
in Figures 1 and 2 as filled stars). It is clear that these objects --- comprising $\sim$20\% 
of the full sample --- tend to fall above the main correlation in the $L_{[OIII]}$  vs.  
$L_{70\mu m}$ plot, 
but lie closer to the main correlation in the  $L_{[OIII]}$ vs.$L_{24\mu m}$ plot; 
the optical starburst 
objects have their 70$\mu$m luminosities enhanced by up to an order of magnitude with respect to 
those without clear signs of star formation activity. We can quantify this difference in 
terms of the vertical displacements of the points relative to the regression line in Figure 
2. Using a Kolmogorov-Smirnoff two sample test to compare the distributions of 
vertical displacements, we find that we can reject the null hypothesis that the starburst 
and non-starburst sub-samples are drawn from the same parent distribution at the 
P=0.005 level of significance ($n_1=39$, $n_2=8$, one-tailed test). This result is further 
reinforced if we consider the supplementary sample of all the radio-loud AGN from 
outside our sample known to show signs of optical star formation activity (open 
stars in Figures 1 and 2). Note that the presence of significant starburst heating in a 
subset of our sample is consistent with recent results obtained for radio-quiet quasars 
based on mid-IR detection of PAH features \citep{schweizer06} and radio continuum data \citep{barthel06}.

As an alternative to optical continuum properties, the MFIR colors may also be 
used to investigate whether star formation --- in this case heavily obscured star formation 
--- is important in the target galaxies. Figure 3 shows the distribution of F(70)/F(24) 
colors for the full 2Jy sample, as well as the starburst and non-starburst sub-samples. It 
is striking that many of the objects with optical star formation activity have relatively 
``cool'' colors ($F(70)/F(24)>3.5$) consistent with those of starburst galaxies in general. 
On the other hand, most of the objects without clearly identified optical star formation 
activity have warmer colors ($F(70)/F(24)<3.5$); using a two sample Kolmogorov-Smirnoff 
test we find that this difference is significant at the P=0.005 level ($n_1=39$, 
$n_2=8$, one-tailed test). This reinforces the view that the reason for the large scatter in the 
70$\mu$m correlation is a group of objects that have enhanced 70$\mu$m luminosities due to a 
contribution from starburst heating. 

\section{Discussion and conclusions}

Given the similarities between Figures 1 and 2, as well as the measured slopes of the 
correlations, it is likely that the dominant heating mechanism for the dust emitting at 
both 24$\mu$m and 70$\mu$m is AGN illumination, with starburst heating contributing 
significantly at 70$\mu$m only in the minority of objects with independent evidence for 
recent star formation activity. However, there is also evidence for a loose correlation 
between starburst and AGN activity, in the sense that the most luminous starbursts 
($L_{70\mu m} > 10^{25}$~W Hz$^{-1}$) are only found in the objects with the most powerful AGN activity 
($L_{[OIII]} > 10^{36}$~W). 

It also is notable that slopes determined for the main correlations shown in Figures 1 and 2 are in 
good agreement with the predictions of simple AGN illumination models involving
photoionization of optically thick clouds  
($L_{MFIR} \propto L_{[OIII]}^{0.83\pm0.1}$: Tadhunter et al. 1998), provided that the relative covering factors of narrow emission line 
region ($f_{NLR}$), the mid-IR emitting dust structure ($f_{MIR}$), and the far-IR emitting dust 
structure ($f_{FIR}$) do not change substantially with luminosity. 
In this context it is interesting to consider whether AGN illumination is energetically feasible.
We find that, in order to 
explain the normalisations of the main correlations apparent in Figures 1 and 2, we 
require  $f_{MIR}/f_{NLR} \sim 12$ and $f_{MIR}/f_{NLR} \sim 6$\footnote{For the purposes
of this calculation we make the following assumptions: case B recombination for
an electron temperature of $T_e = 15,000$~K; $[OIII]\lambda5007/H\beta = 12$ and
a mean ionizing photon energy of $\langle hv \rangle =38.7$~eV (see Robinson et al. 1987); and a ratio of ionizing luminosity to bolometric
luminosity of $L_{ION}/L_{BOL} = 0.32$ (see Elvis et al. 2004). The mid-IR and far-IR luminosities
have been integrated over the wavelength ranges 3 -- 30$\mu$m and 30 -- 100$\mu$m respectively
assuming a spectral index of $\alpha = 0.7$ ($F \propto \nu^{-\alpha}$), estimated from the 
mean F(70)/F(24) 
flux ratio for the sample as a whole.
Note that
no assumptions have been made about the detailed radial distribition of dust. We
simply assume that the dust is distributed in such way that it produces the observed
MFIR SEDs by AGN illumination.}. This implies that the 
MFIR emitting dust 
structures cover a substantially larger fraction of the sky than the narrow emission 
line region (NLR). Given that the covering factor of the NLR is typically a few percent, 
the dust structures are likely to cover $\sim$20 -- 70\% of the sky as seen by the AGN. This is entirely 
feasible if the dust is associated with the central obscuring tori required by the unified 
schemes for powerful radio sources \citep{barthel89}, or with the kpc-scale dust lanes visible in high 
resolution images of some radio galaxies \citep{dekoff00}.

On the basis of our results it is clear that powerful, radio-loud AGN are not always
accompanied by major contemporaneous starburst episodes: considering both the MFIR 
colours and the optical continuum spectra we estimate that the proportion of radio galaxies 
in our sample with significant recent starburst activity (optically obscured or otherwise) 
falls in the range 20 -- 30\%. We hypothesise that the presence of a major starburst 
component, as revealed by enhanced 70$\mu$m emission, is related to the mode of 
triggering of the AGN and radio jets. For example, it is plausible that the radio galaxies 
with starburst components are triggered relatively close to the peak of starburst activity 
in major, gas-rich galaxy mergers, whereas those lacking significant starbursts are 
triggered later in the merger sequence (e.g. Tadhunter et al. 2005), by relatively minor accretion events, or by 
cooling flows \citep{bremer97}. This would be consistent with the observed morphological
and kinematical diversity of the population of powerful radio galaxies (Heckman et al. 1986;
Tadhunter, Fosbury \& Quinn 1989; Baum, Heckman \& van Breugel 1992).
It will be possible to test these ideas in future by using deep optical 
imaging observations to relate the interaction status and environments of the host 
galaxies to the MFIR properties revealed by Spitzer.

\acknowledgments We thank the anonymous referee for
useful comments. This work is based on observations obtained with the 
{\it Spitzer Space Telescope}, which is operated by the Jet Propulsion 
Laboratory, California Institute of Technology under NASA contract 1407.
DD, JH and KI acknolwledge support from PPARC.



{\it Facilities:} \facility{Spitzer (MIPS)}

\clearpage

\begin{figure}
\epsscale{.80}
\plotone{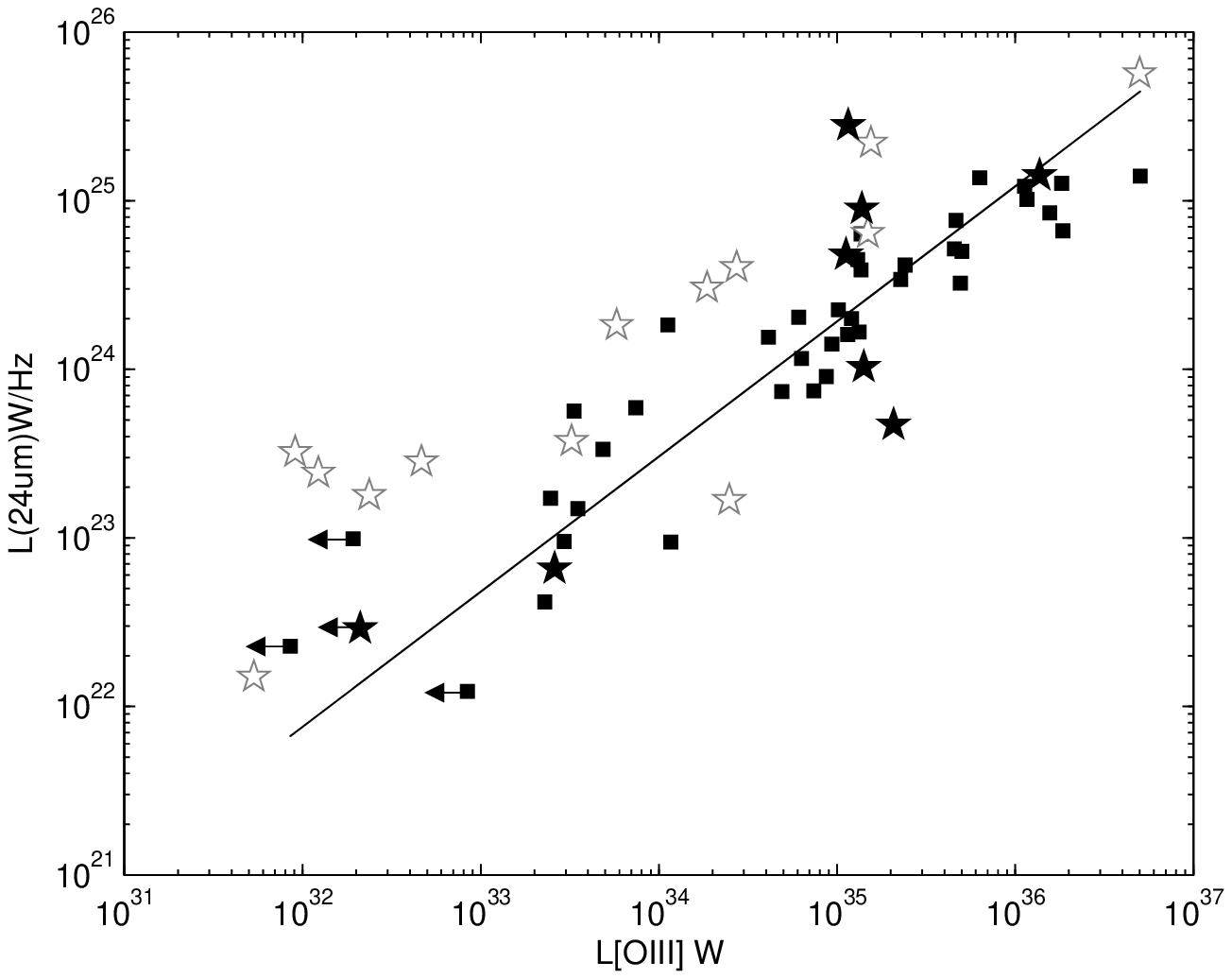}
\caption{24$\mu$m monochromatic luminosity plotted against 
Galactic reddening corrected [OIII]$\lambda$5007 emission line luminosity for 
the 2Jy sample. The filled stars and  squares show respectively 
objects with and without clear spectroscopic evidence for recent star formation 
activity at optical wavelengths, while the open stars represent a supplementary 
sample of radio-loud AGN with clear optical evidence for recent star formation activity 
taken from the literature; arrows represent upper limits. The line shows a 
linear least squares fit to the points, calculated as the bisector of the linear 
regression of x on y and y on x for the full sample of objects
with $z>0.06$.} \label{fig1}
\end{figure}

\clearpage

\begin{figure}
\epsscale{.80}
\plotone{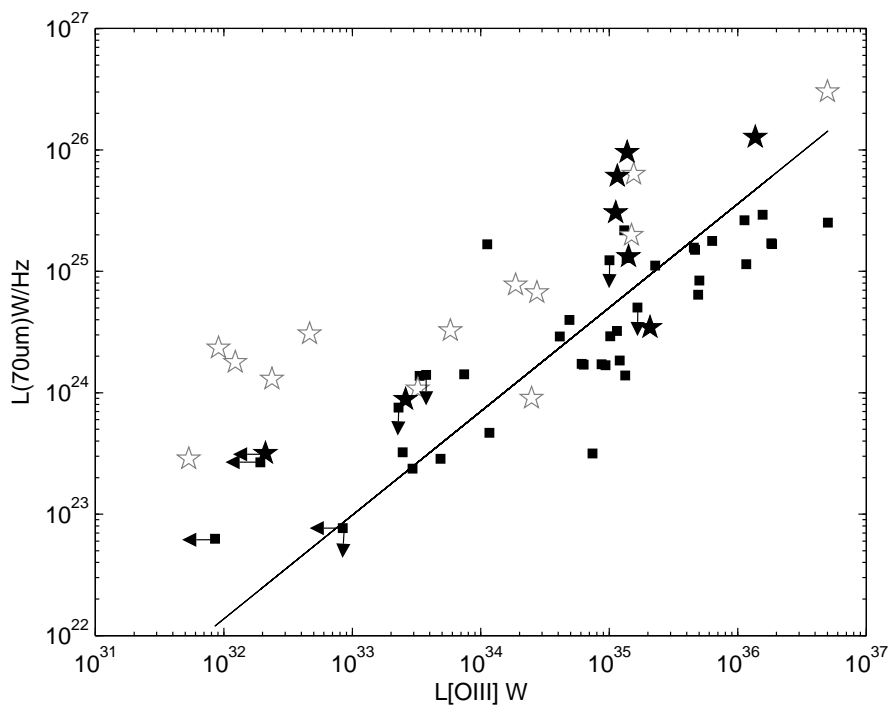}
\caption{70$\mu$m monochromatic luminosity plotted against 
Galactic reddening corrected [OIII]$\lambda$5007 emission line luminosity  
for the 2Jy sample. The symbols are the same as for Figure 1.   
Note that some of the 
objects identified as belonging to the sub-sample without optical evidence for 
star formation activity (black squares), but falling well above (0.3dex, $\sim$3$\sigma$
for an uncertainty of 30\%) the regression 
line, are classified as broad-line AGN. For such objects the emission 
from the AGN swamps the optical spectrum, and it would be difficult to deduce 
the presence of recent star formation activity even if present.}\label{fig2}
\end{figure}

\clearpage

\begin{figure}
\epsscale{.80}
\plotone{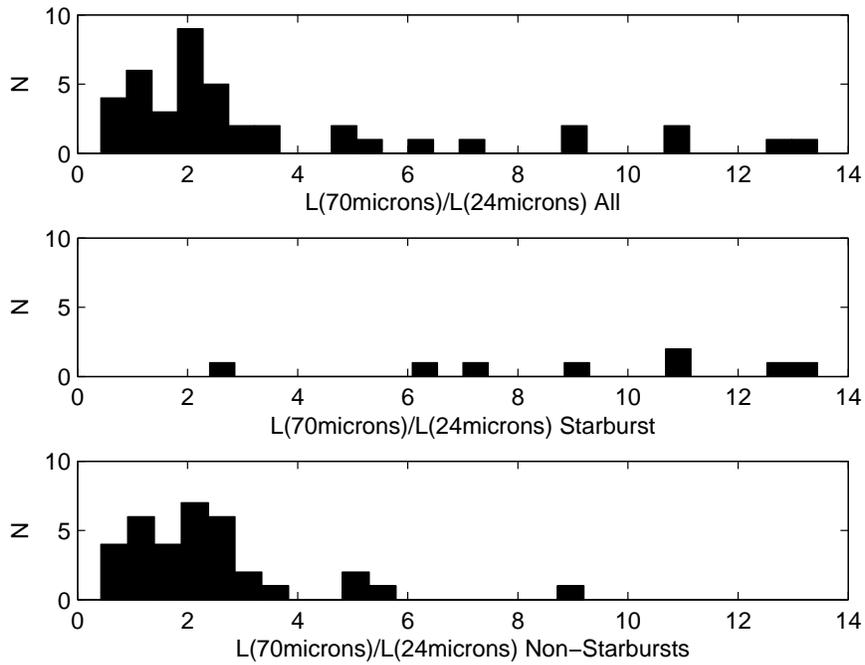}
\caption{Histograms showing the distributions of the measured 70$\mu$m/24$\mu$m 
luminosity ratios for the full 2Jy sample (top), the sub-sample with clear 
spectroscopic evidence for optical starburst activity (middle), and the 
sub-sample without clear evidence for optical starburst activity (bottom).}\label{fig3}
\end{figure}

\clearpage 
\begin{deluxetable}{lllll}
\tabletypesize{\scriptsize}
\tablecaption{Correlation analysis for the 2Jy sample. The second column
gives the sample size, the third column the Spearman rank correlation coefficient 
($z > 0.06$ sub-sample),
the fourth column the significance level for a one-tailed test (i.e. probability that
the quantities are mutually independent), and
the fifth column the standard deviation of the vertical displacements of the MFIR
luminosities for the full sample relative to the best 
fitting regression lines (see Figures 1 and 2).
All the statistics have been calculated using log quantities.}
\tablewidth{0pt}
\tablehead{
\colhead{Correlation} &\colhead{N} &\colhead{$r_s$} 
&\colhead{P} &\colhead{Scatter (dex)}}
\startdata
Including starburst objects: & & & & \\
P$_{24}$ vs. L$_{[OIII]}$ &39 &0.83 &$<$0.0005 &0.40 \\
P$_{70}$ vs. L$_{[OIII]}$ &39 &0.67 &$<$0.0005 &0.57 \\
Without starburst objects:& & & & \\
P$_{24}$ vs. L$_{[OIII]}$ &32 &0.85 &$<$0.0005 &0.34 \\
P$_{70}$ vs. L$_{[OIII]}$ &32 &0.77 &$<$0.0005 &0.51 \\
\enddata
\end{deluxetable}
\end{document}